\documentclass[12pt,preprint]{aastex}
 \usepackage{ifpdf}
 \usepackage{epsfig}
  \usepackage[latin1]{inputenc}
 \usepackage{amsfonts}
 \usepackage{amsthm}
 \usepackage{amssymb, latexsym, amsmath}
 \usepackage{amsbsy}
\usepackage{amstext}
\usepackage{amsopn}
\usepackage{amsgen}
\usepackage[dvips]{color}
 \usepackage{graphicx}
\usepackage{xcolor}
\definecolor{rkka}{RGB}{219,66,32}
\newcommand{\eb}{\begin{equation}}
\newcommand{\ee}{\end{equation}}
\shorttitle{Orbital elements of long-period binaries}
\shortauthors{Makarov}
\begin{document}
\title{Two-epoch orbit estimation for wide binaries resolved in Hipparcos and Gaia}
\author{Valeri V. Makarov}
\email{valeri.makarov@navy.mil} 
\affil{US Naval Observatory, 3450 Massachusetts Ave NW, Washington DC 20392-5420, USA}

\date{Accepted . Received ; in original form }

\label{firstpage}
\begin{abstract}
The Hipparcos catalog and its Double and Multiple System Annex (DMSA) lists 4099 components
with individual proper motions and coordinates on the epoch 1991.25. Many of these long-period binary stars
are also present in the Gaia Data Release 2 (DR2). Using the available relative positions and
proper motions separated
by 24.25 years, the equations of relative orbital motion can be solved for the two
epoch eccentric anomalies, orbital period, and eccentricity. This method employs elimination
of the linear Thiele-Innes unknowns and nonlinear optimization of the remaining condition equations.
The quality of these
solutions is compromised by the insufficient condition and modest precision of the
Hipparcos astrometric data, as revealed by Monte Carlo simulations with artificially perturbed
data points. The presence of multiple systems and optical pairs can also perturb the results.
Limited experiments with artificial data indicate that useful estimates can be obtained with a 25-year
epoch difference for wide binaries with orbital periods up to $\sim 500$ years. The prospects of this
method dramatically improve with the proposed next-generation space astrometry missions such as Gaia-NIR
and Theia, especially when additional conditions are included from astrometric or spectroscopic measurements.
An  ancillary catalog of cross-identification and astrometric information for 1295 double star pairs cross-matched
in Gaia DR2 and Hipparcos is published.

\end{abstract}


\section{Introduction}
The Gaia astrometric and photometric mission \citep{pru,bra} provided a trove of new data for billions of celestial
objects, including the best studied nearby and bright stars, which are important for navigation and space
awareness applications. The multiplicity fraction of nearby solar-type stars is 0.46 \citep{tok14}, and the
empirically fitted log-normal distribution of orbital periods suggests that half of binary pairs have
periods longer than 100 yr, while 16\% of them have periods between 100 and 1000 yr. Such binary systems
at 10 pc distance have upper-bound separations in the range $3\arcsec$ -- $15\arcsec$ and should be easily
resolved by both Hipparcos \citep{esa} and Gaia. The rate of spectroscopic binary and multiple systems
among the bright red giants is estimated to be 0.46 at the 75\% confidence level \citep{maku}. Magnitude-limited
samples of navigation stars are biased toward red giants and early-type dwarfs, owing to their higher
luminosity, and they probe larger volumes of space. We should expect no less than 7\% of all bright stars
to have physical companions with periods between 100 and 1000 yr. Even though the brightest targets of this
category have been observed by double star researchers for over a century, only a small fraction of them
have well-defined visual orbits. 

Orbit determination for visual binaries is a hard and laborious process that is also slow for systems with
long periods ($>100$ yr). Even when both components are sufficiently bright and the separation is greater
than the typical seeing or angular resolution, observers have to collect multiple position measurements
at different phases of the orbit, which requires a continuous observing campaign. High-eccentricity
systems are especially problematic because they spend most of the time closer to apoastron where little
changes can be measured. In the era of global space astrometry and large etendue ground-based surveys,
one would like to utilize the available astrometric data characterized by a limited time span and number
of observational epochs. The highest precision, which is essential for orbital work, is provided by
the Hipparcos space mission \citep{lin,per} and Gaia, currently available as Data Release 2. Both mission
durations are short enough to consider the mean astrometry data as single epoch points of long-period
systems. Is orbit estimation possible with these limited data? The goal of this paper is to demonstrate
that it is possible in principle, and to demonstrate the application of the new two-epoch technique
to real space astrometry data. The idea is similar to decoupling the 7 orbital parameters into groups
that are almost or explicitly independent of each other and estimating these groups in steps \citep{des}.

Orbital motion perturbs the observed proper motion of binary components, and this perturbation has been
used to detect new binaries from astrometric measurements \citep{wil, maka, ker}, or to back-engineer
the distribution of important orbital parameters from an observed sample distribution of astrometric
parameters \citep{tokk}. In the proposed method, the relative proper motions are directly used to fit
orbits. This novelty will prove especially advantageous when a follow-up space astrometry mission to
the currently operating Gaia, such as the proposed Gaia-NIR \citep{mca} or Theia \citep{the} delivers
sub-1 mas astrometry for millions of binary systems.
The method is described in \S\ref{sol.sec}. It is applied to a sample of visual double stars collected
from the Hipparcos and Gaia DR2 catalogs (\S\ref{sam.sec}), but extensive Monte Carlo simulations with
randomly perturbed input epochs revealed insufficient performance with four free orbital parameters (\S\ref{perf.sec}).
Useful improvements can be implemented by reducing the number of degrees of freedom with an additional
hard constraint derived from the Kepler's equation and by obtaining solutions for a grid of fixed
first-epoch eccentric anomaly values (\S\ref{impr.sec}). Fixed eccentricity solutions
for 151 pairs (\S\ref{fix.sec}) based on Hipparcos Gaia data turn out to be marginally successful.
The limitations of the technique come from the modest precision of the available data and the presence
of hierarchical multiples and optical pairs. The prospects with future space astrometry missions
are briefly discussed in \S\ref{dis.sec}.

\section{Two-epoch solution}
\label{sol.sec}

The apparent motion of a binary in the plane of
celestial projection is described by \citep{hei}:
\begin{eqnarray}
x &=&A(\cos E-e)+F\sqrt{1-e^2}\sin E \\ \nonumber
y &=&B(\cos E-e)+G\sqrt{1-e^2}\sin E 
\label{coo.eq}
\end{eqnarray}
where $x$ and $y$ are the relative angular tangential coordinates,
and $E$ is the eccentric anomaly related to the mean anomaly $M$ by
Kepler's equation
\eb
M=2\pi\frac{T-T_0}{P}=E-e\sin E.
\ee
The Thiele-Innes constants are related to the remaining orbital
elements by
\begin{eqnarray}
A &=&a(\cos \omega\cos \Omega -\sin \omega \sin \Omega \cos i) \\ \nonumber
B &=&a(\cos \omega\sin \Omega +\sin \omega \cos \Omega \cos i) \\ \nonumber
F &=&a(-\sin \omega\cos \Omega -\cos \omega \sin \Omega \cos i) \\ \nonumber
G &=&a(-\sin \omega\sin \Omega +\cos \omega \cos \Omega \cos i), 
\end{eqnarray}
where $a$ is the angular semimajor axis, $\omega$ is the periastron longitude,
$\Omega$ is the node, and $i$ is the orbit inclination ($i=90\degr$ is an edge-on orbit).

The angular elements $\omega$ and $\Omega$ refer to a specific celestial coordinate system. Traditionally,
the equatorial system is used with a fixed vernal equinox. In this case, $x=\Delta \alpha \cos\delta$,
$y=\Delta\delta$, where $\alpha$ and $\delta$ are the right ascension and declination of a given
point on the sky. The actual projected trajectory of a binary component differs from a perfect ellipse
because of the curvature of the celestial sphere, but this very small deviation can be ignored. All the
orbital parameters are considered to be constant in time except $E$, which is a good approximation
when the effects of perspective acceleration, light travel time, parallactic angle, Galactic
tidal perturbations, and possible
dynamical evolution or relativistic effects can be neglected. It is straightforward to differentiate
the astrometric equations in time in order to link the orbital parameters with the observable
proper motions:
\begin{eqnarray}
\dot x &=&(-A\sin E+F\sqrt{1-e^2}\cos E)\dot E \\ \nonumber
\dot y &=&(-B\sin E+G\sqrt{1-e^2}\cos E)\dot E 
\label{dif.eq}
\end{eqnarray}
where $\dot E=n/(1-e\cos E)$, and $n=2\pi/P$ is the mean motion. In the equatorial celestial system,
the derivatives are equal to the components of instantaneous proper motion caused by the orbital motion, i.e.,
$\dot x=\mu_{\alpha *}$, $\dot x=\mu_{\delta}$, if $a_a$ is in mas and $T$ is in years. These
equations are valid for each of the binary components or for the observed photocenter of unresolved
pairs, in which case a smaller photocenter excursion $a_a$ should be used instead of the actual
projected semimajor axis $a$.

This investigation concerns well-separated, long-period, resolved pairs with separate positions and
proper motions determined in Hipparcos and Gaia. The orbital period should be
longer than 24.25 years, which is the time difference between the epochs of observation. The upper
limit is not well defined, because it depends on the distance, eccentricity, and geometric
configuration, but it is probably on the order of 500 -- 1000 years.
The components' proper motions include the
barycentric proper motion, which cannot be determined, because the mass ratio of the companions is
not {\it a priori} known. The unknown barycenter position and motion can be eliminated if we consider
the relative values, i.e., the difference between $x$, $y$, $\dot x$, and $\dot y$ of the secondary
and the primary components. Equations 1 and 4 remain valid for the differential
data, with $a$ corresponding to the total mass of the system. A set of relative coordinates and
proper motions from an astrometric catalog provides four condition equations, which can be written
in a matrix form:
\eb
\bf{Q}\cdot \boldsymbol{p} = \boldsymbol{d},
\ee
where vector $\boldsymbol{d}=(x,y,\dot x, \dot y)^T$ comprises observational data, $\boldsymbol{p}=(A,F,B,G)^T$
comprises the unknown Thiele-Innes constants, and $\bf{Q}$ is a 4 by 4 matrix comprising functions of $e$,
$E$, and $\dot E$, which are readily derived from Eqs. 1 and 4. Inverting this
equation,
\eb
\boldsymbol{p} = \bf{Q}^{-1}\cdot \boldsymbol{d}.
\ee
The rank of this system with 7 unknowns is 4, so it is undetermined. However, if we have data from two
independent epochs (1 and 2), the number of unknowns is explicitly 8 (with $E$ splitting into $E_1$
and $E_2$), and the system can be numerically solved. This can be done via eliminating the linear
part of the equation, which concerns $\boldsymbol{p}$, e.g.,
\eb
\mathbf{Q}_2\bf{Q}_1^{-1}\cdot \boldsymbol{d_1} = \boldsymbol{d_2}.
\label{eq.eq}
\ee
Alternatively, the observations at epoch 1 could be kept in the right-hand part and epoch 2 data used
to eliminate the linear unknowns, which is equivalent to swapping indices 1 and 2 in this equation. 
The matrix $\bf{S}=\bf{Q}_2\bf{Q}_1^{-1}$ is, specifically,
\eb
\label{s.eq}
\bf{S}=
  \left[ {\begin{array}{cccc}
   \frac{\cos(E_1-E_2)-e\,\cos E_1}{1-e\,\cos E_1} & 0 & 
   \frac{\sin(E_2-E_1)-e(\sin E_2-\sin E_1)}{n} & 0 \\
   0 & \frac{\cos(E_1-E_2)-e\,\cos E_1}{1-e\,\cos E_1} & 
   0 & \frac{\sin(E_2-E_1)-e(\sin E_2-\sin E_1)}{n} \\
   \frac{n\,\sin(E_1-E_2)}{(1-e\,\cos E_1)(1-e\,\cos E_2)} & 0 &
   \frac{\cos(E_1-E_2)-e\,\cos E_2}{1-e\,\cos E_2} & 0 \\
   0 & \frac{n\,\sin(E_1-E_2)}{(1-e\,\cos E_1)(1-e\,\cos E_2)} &
   0 & \frac{\cos(E_1-E_2)-e\,\cos E_2}{1-e\,\cos E_2} \\
  \end{array} } \right]
\ee

This system of equations can be solved numerically by the nonlinear least squares method, minimizing the
square norm $\| \boldsymbol{d_2}-\bf{S}\cdot \boldsymbol{d_1} \|^2$ using any of the standard nonlinear optimization
techniques. Alternatively, a 1-norm solution can be implemented minimizing the sum of the absolute values of
these residuals.

\section{Preparing the sample}
\label{sam.sec}

We start with the Hipparcos DMSA Component Solution catalog downloaded from the Vizier database. This data set includes
24588 records with 22 fields each, a mixture of string and numerical values. Each record corresponds to a resolved
component of double or multiple systems. Most of the components in this collection have constrained solutions indicated
with a flag of F in the first field. The photomultiplier detector of the main instrument with a sensitive area spot
of $\sim 38\arcsec$ diameter was tracking one star at a time, but any other source within that area contributed to
the combined light, which was modulated on a grid of slits. Therefore, most double stars with separations $20\arcsec$
or less could not be separated at the level of the detected photons, and the regular 5-parameter adjustment could not
be applied. The presence of another star shifts the phase and changes the amplitudes of the two recorded harmonics of
the modulated signal. To disentangle this signal with 10 astrometric unknowns is difficult---and often impossible without
sufficiently accurate prior knowledge of the relative position and brightness of the components. As a result, many
double star components were ``fixed" to have the same parallax and proper motion, reducing the number of astrometric unknowns to 7.

We find only 4099 entries that were not constrained such that they may have independently determined parallaxes and proper motions.
The next step is find their counterparts in the Gaia DR2 catalog. The mean positions at 1991.25 were transferred to the
Gaia DR2 epoch, which is 2015.5, using Hipparcos proper motions from the DMSA. A cone search with a $2\arcsec$ radius
resulted in 4128 tentative matches. Some of the Hipparcos pairs are completely missing in Gaia or are present without
measured proper motions. This mostly concerns doubles with small separations\footnote{As a reminder, the hard lower bound
on separation in Hipparcos is $0.1\arcsec$.}. Apparently, the Gaia pipeline could not handle tight doubles and abandoned
these targets. At larger separations, only the secondaries are sometimes missing, e.g., the B component of HIP 375
at $\rho=22.72\arcsec$. There is a star in Gaia that may be the actual counterpart but with a greatly different proper
motion. We suspect an optical pair in this case with a grossly incorrect data in Hipparcos.
About one-tenth of the initial sample of components (442) could not be reliably matched with Gaia. They may be real
omissions in Gaia for complicated close doubles or gross errors in Hipparcos for optical pairs at large separations.

Many Hipparcos components get matched with more than one Gaia counterpart. This happens, for example, for HIP 110 at
$\rho=1.2\arcsec$ where each of the components are matched with either part of a resolved Gaia pair. To avoid the confusion
at small separations caused by the coarse cone search radius, a merit function {\it merit = 2$\times$AbsDistance + $|$Gmag-Hp$|$ }
was applied to all cases of multiple identification. The resulting sample includes 3342 uniquely cross-matched Hipparcos
components. Some of them are cross-matched to the same Gaia entry whenever Gaia failed to resolve the system. Eliminating those,
as well as 3 bogus triple systems, results in a set of 3011 components: 1698 of them are A components, and 1313 are B components.
Pairs with only one of the components in Gaia are useless for this analysis, so the pairs have to be identified and
coupled from scratch, because the HIP numbers can be different within a pair (example: HIP 71 (A) and 70 (B)), so they cannot 
be relied upon. This coupling results in a catalog of reliably cross-matched 2590 components of 1295 doubles,
which is separately published online. Each star has independently determined positions and proper motions from
Hipparcos and Gaia.

\section{Computational performance and confidence intervals}
\label{perf.sec}
Equations \ref{eq.eq} can be numerically solved by a variety of nonlinear optimization methods to yield estimates of
$E_1$, $E_2$, $e$, and $n$. Although the number of equations equals the number of formal unknowns, the vectors
$\mathbf{S}\cdot \boldsymbol{d_1}$ and  $\boldsymbol{d_2}$ are never equal because both parts include random errors.
The data vectors are, specifically, $\boldsymbol{d_i} = \{\Delta\alpha_i \cos \delta_i,\;
\Delta\delta_i,\; \Delta\mu_{\alpha,i} \cos \delta_i,\; \Delta\mu_{\delta,i}\}'$, $i=1,2$, where all deltas denote the
difference ``component B minus component A", and $i$ is the index of observation epoch.
An optimal solution that minimizes a certain metric of the difference should be sought. One investigated option is to use
nonlinear least-squares method to minimize the Euclidian norm of the difference. The other possibility is to use a 1-norm
metric that would find the smallest sum of absolute values of the difference vector elements. The latter approach may be
more stable in the presence of large outliers (flukes), to which a least-squares method is sensitive. 

Early trial calculations with both techniques revealed that a vast majority of 1295 selected pairs did not provide
reasonable solutions. The main reason is the modest precision of Hipparcos data from the special component solution.
Long-period binaries with periods greater than 100 yr may have small proper motion changes on the timescale of 25 yr,
unless they are close to us. The situation is reminiscent to trying to measure a small distance between two points
by using a measure tape with a precision similar to the distance. The main manifestation of inadequate precision is
the high rate of ``runaway" solutions with unrealistically high eccentricity (close to 1) and long orbital periods
exceeding 1000 -- 10000 yr. A random observational error makes the proper motion difference between the two epoch
much larger than the true value, and the nonlinear optimization tries to capture this change in proper motion fitting
a very elongated ellipse and placing both epochs close to the periastron. Attempting to mitigate this objective
difficulty, I removed all data with signal-to-noise ratio below 3 in Hipparcos proper motion difference
between the components of each pair, i.e., $\Delta\mu
< 3\;\sigma_{\Delta\mu}$. The sample was decimated to a meager 472 components of 236 pairs.

\begin{figure}[htbp]
\epsscale{0.75}
  \centering
  \includegraphics[angle=0,width=0.78\textwidth]{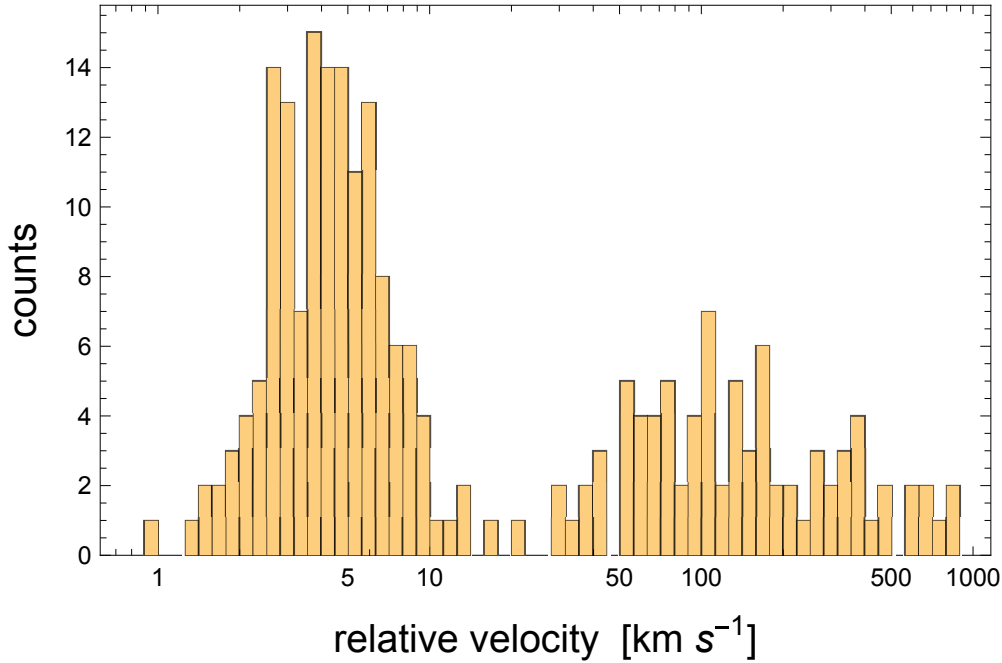}
\hspace{2pc}
\caption{Histogram of estimated relative orbital velocity differences for 236 resolved double stars with independently
determined Hipparcos and Gaia DR2 astrometric parameters.}
\label{d_v.fig}
\end{figure}

The unavoidable difficulty with this selection is that Hipparcos provides relatively small proper motion errors
only for widely separated pairs, which tend to be optical. This category can be characterized from significantly
different parallaxes in Gaia, but a spot check revealed that Gaia parallaxes for double stars cannot always be
trusted. An estimated projected orbital velocity was used instead to remove the most obvious optical pairs:
\eb
\Delta v = 4.74\cdot \Delta\mu/\varpi,
\ee
where proper motion is in mas yr$^{-1}$, parallax $\varpi$ from Gaia is in mas, and $\Delta v$ is in km s$^{-1}$.
Fig. \ref{d_v.fig} shows the histogram of velocity changes thus computed (note the logarithmic scale of the
abscissa axis). The distribution is clearly bimodal, with a gap separating two populations. The large $\Delta v$
pairs are mostly optical, with possibly a small addition of nearby short-period binaries. All pairs with $\Delta v>25$
km s$^{-1}$ were removed, leaving 151 pairs of 302 components.

Using the 2-norm option (i.e., nonlinear least-squares optimization) with weights constructed from the formal errors
in Hipparcos and Gaia for each of the four condition equations, a unique set of parameters $\{ E_1, E_2, e, n \}$
can be computed. These solutions are often of low value, however, due to the insufficient signal-to-noise ratios.
To estimate the uncertainty of the results, extensive Monte Carlo simulations were performed for all 151 pairs using
parallel computing in Mathematica. The observed data points $\boldsymbol{d}_1$ and  $\boldsymbol{d}_2$ were
additively perturbed by Gaussian-distributed random variables with a zero mean and standard deviations equal to
their combined formal errors. For each trial, a set of 101 perturbed data vectors was generated, and a complete
optimization solution obtained for each realization by the Random Search algorithm. The latter was found to perform somewhat better
than the more traditional gradient or simulated annealing techniques because of multiple local minima of the
merit function in the examined parameter space. The quantiles of the emerging distributions of eccentricity and
orbital period $P_{\rm orb}=2\pi/n$ provide confidence intervals of the estimated two-point solutions.

\begin{figure}[htbp]
\epsscale{0.75}
  \centering
  \includegraphics[angle=0,width=0.78\textwidth]{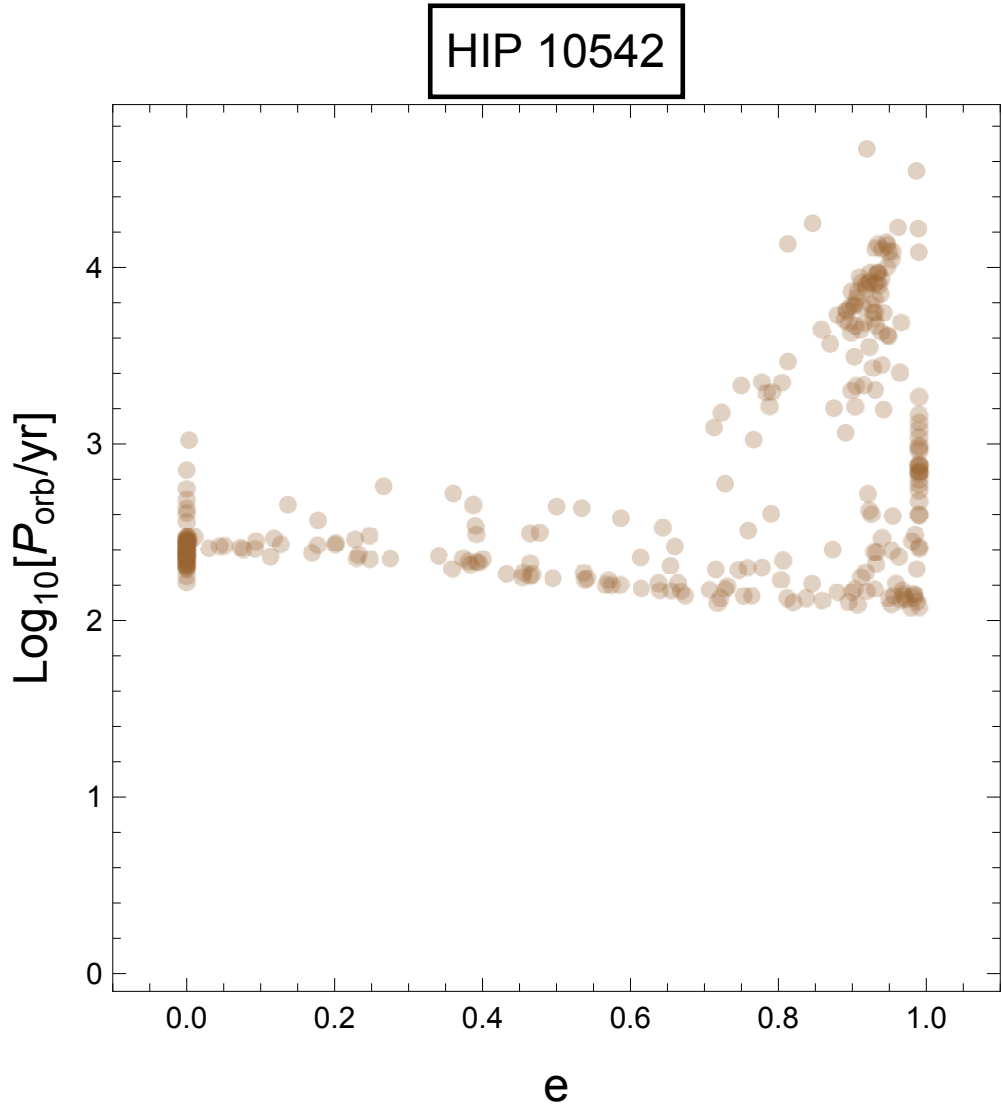}
\hspace{2pc}
\caption{Results of 301 Monte-Carlo simulations of two-point orbit solutions for the physical
binary HIP 10542 resolved in Hipparcos and Gaia DR2.}
\label{dis.fig}
\end{figure}

Fig. \ref{dis.fig} shows the distribution of eccentricity and $\log P$ values for one specific system HIP 10542 =
WDS $02158-1814$. It is a known visual double with estimated $e=0.2$ and $P=330$ yr \citep{izm} extensively
studied by speckle imaging \citep{ho00,mas04,ho06,tok10,tok15}. Although the Monte Carlo trials do cluster
around the known period at small eccentricity\footnote{Our estimation is more consistent with the earlier
orbits for the system, e.g., $P=225$ yr from \citep{so99}.}, the eccentricity itself is not well constrained, with clumps
at $e=0$, $e=1$ and a separate group of bogus solutions with high $e$ and unreasonably long periods $\log P>3$.
This illustrates a rather common issue with the two-point orbit estimation. Some combinations of randomly perturbed
data yield runaway solutions with eccentricity values often piling up to the upper bound and unrealistically
long periods. The majority of other stars show a similar pile-up at $e=0.99$, which was a hard constraint on the
optimization. A poor condition of eccentricity is a known generic problem for astrometric orbital evaluation,
which was encountered, for example, in Hipparcos-only estimation of short-period binaries \citep{goma}.

\section{Possible improvements and experiments with artificial data}
\label{impr.sec}
The basic algorithm of two-point orbit estimation described above seeks a solution for four orbital parameters
$\{E_1,E_2,e,n\}$ after explicit elimination of the Thiele-Innes coefficients using the Gaia data point. However,
this set of unknowns is redundant, because there are only 7 independent orbital elements. One can eliminate
this redundancy by introducing one additional constraint from the Kepler's equation:
\eb
e\,(\sin E_2 - \sin E_1) = E_2 - E_1 - n\,\Delta t,
\label{kep.eq}
\ee
where $\Delta t$ is the epoch difference, which equals to 24.25 years for the Hipparcos and Gaia DR2 data sets. 
Technically, this can be implemented as a constrained nonlinear optimization problem, making the solution
somewhat more computer-intensive. At first glance, this should improve the performance, because the number of
degrees of freedom is reduced by one. In reality, as my numerical experiments with artificial noiseless data
revealed, it only helps to achieve a more accurate solution in the case of negligibly small noise. No significant
improvement has been found when applied to the real data in hand. As a possible explanation, we note that
statistical flukes or large errors in combination with the additional hard constraint linking the eccentric anomalies (Eq. \ref{kep.eq})
can perturb the parameters of significance $e$ and $n$ even more.

Another possible improvement concerns the possibility of more stable solutions on a grid of eccentric anomalies
at first epoch, $E_1$. Fixing this less essential parameter, rather than orbital eccentricity, may provide insight into
the overall behavior and condition of the optimization process, as well as help to identify the global minimum of the merit
function. To test this hypothesis, I generated synthetic relative positions and proper motions at the relevant epochs
1991.25 and 2015.5 for a binary systems with $i=75\degr$, $\omega=120\degr$, $\Omega=200\degr$, $T_0=1900$, $P_{\rm orb}=500$
yr, $e=0.4$, $M_{\rm tot}=2 M_{\sun}$ at 100 pc from the Sun. The simulated data sets were processed with the constrained
method using different nonlinear optimization techniques: random seeded search, Nelder-Mead, differential evolution,
and simulated annealing. The random search solution with 400 search points produced a very good
result, with $e=0.387$, $n=0.011$ rad$/$yr. The other algorithms fail to converge to the true orbit, leaving significant
errors in $e$ and $n$ despite the anomaly condition Eq. \ref{kep.eq} being rigorously satisfied. The reason for this is
the complex structure of the merit function within the allowed multi-dimensional parameter space taking numerous minima,
which may only be slightly larger than the true (global) minimum. 

\begin{figure}[htbp]
  \centering
  \includegraphics[angle=0,width=0.48\textwidth]{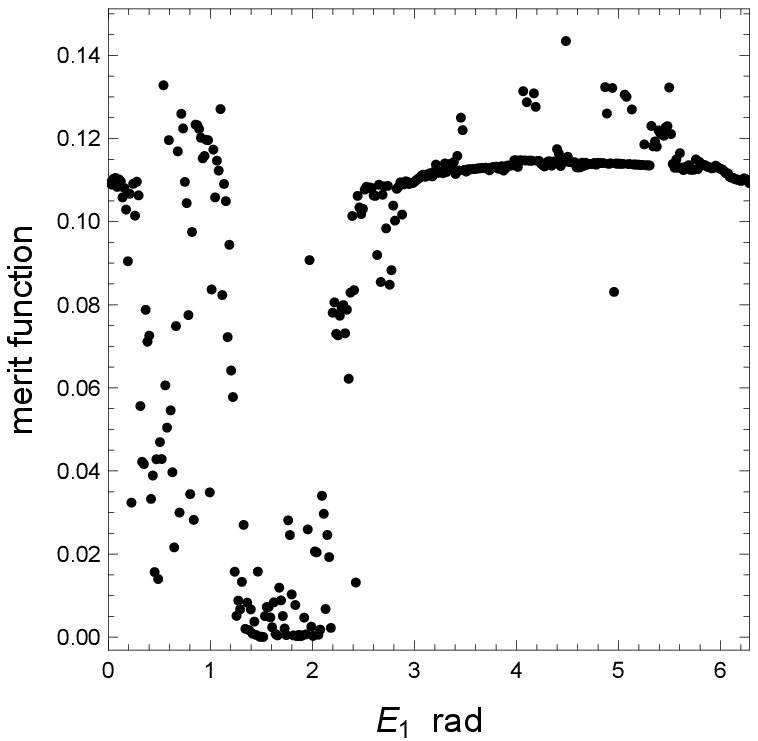}
  \includegraphics[angle=0,width=0.48\textwidth]{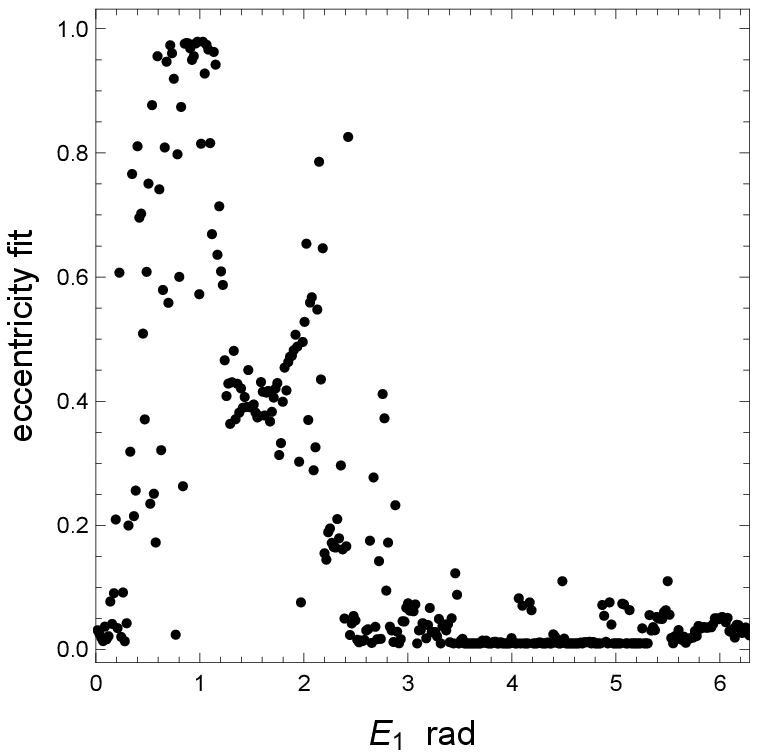}
\hspace{2pc}
\caption{Results of 360 two-point orbit solutions for a synthetic binary system with a period of 500 yr and
noiseless data. The first-epoch eccentric anomaly was fixed on a grid of values covering the interval 0 to $2\pi$.}
\label{e1.fig}
\end{figure}

Mapping the entire parameter space is not feasible, but further insight can be gained by fixing $E_1$, which is more
restrictive than fixing $e$. The random search and Nelder-Mead methods produce practically perfect solutions if $E_1$ value is fixed
at the true value (1.5466 rad). Searching for the global minimum may then be implemented on a sufficiently dense grid
of fixed $E_1$. Using the Nelder-Mead algorithm, which is much faster, I produced orbit solutions for a grid of 360 values
$E_1=0,1,2,\ldots ,359\degr$. Fig. \ref{e1.fig} shows the values of the merit function $||\mathbf{S}\cdot \boldsymbol{d_1} - \boldsymbol{d_2}||$ (left) and the fitted eccentricity $e$ (right) for this grid. Unfortunately, there is a plateau
of $E_1$ values where the merit function is close to 0, which implies that this optimization is intrinsically
uncertain (ill-conditioned). The corresponding estimates of $e$ group around the true value $0.4$, but with some obvious scatter.

\section{Estimating orbital periods at fixed eccentricity and comparison with Izmailov 2019 catalog}
\label{fix.sec}
In view of the poor condition of two-epoch solutions for eccentricity, a more modest objective was set to derive
estimates of orbital periods for 151 binary star systems at three fixed eccentricities: $e=0$, 0.5, and 0.9.
The other three unknowns entering the epoch transformation matrix (Eq. \ref{s.eq}) were constrained in the
random search optimization: $E_1\in [0,2\pi]$, $E_2\in [0,2\pi]$, $n\in [n_{\rm est}/20$, $n_{\rm est}\cdot10]$,
where $n_{\rm est}=\sqrt{0.5\;(\rho/\varpi)^3}$, and both the separation $\rho$ and parallax $\varpi$ were extracted from the
Gaia data. The hard condition on eccentric anomalies was not used in this experiment.
The benchmark $n_{\rm est}$ is a crude estimate of the mean motion. The actual mean motion may deviate
for various reasons, including the total mass being different from the assumed $2\,M_{\sun}$, the semimajor axis being
likely greater than the observed separation due to the projection effect, the semimajor axis sometimes being smaller
than the observed separation due to an eccentric orbit observed closer to the apoastron. The bounds on $n$ are meant
to limit runaway solutions and provide a clear indication of failed optimizations.

The sample median period of $e=0$ solutions is $\sim 630$ yr. In each eccentricity sample, 20 to 23 solutions bump 
into the upper constraint $20\,P_{\rm est}$
resulting in a $> 14$\% rate of runaway solutions. A much higher rate of failed solution emerges from a comparison
with the catalog of orbital parameters by \citet{izm}. I found 83 systems in common with this catalog, 5 of which
are known triple systems where two-point estimates cannot be valid. Only 14 solutions qualify as ``good quality",
whereas 43 systems are a total mismatch. The latter often have unreasonably long periods in excess of 2000 years.
It appears that the method tends to fit the observed position and velocity differences as a high-eccentricity
orbital motion close to the periastron of otherwise a slow orbit.

\begin{figure}[htbp]
  \centering
  \includegraphics[angle=0,width=0.48\textwidth]{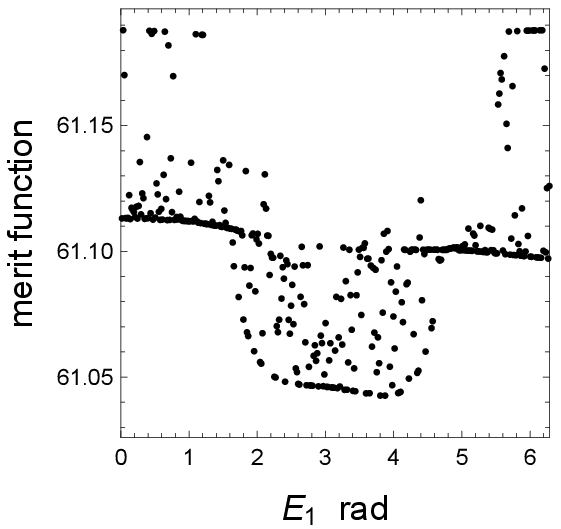}
  \includegraphics[angle=0,width=0.48\textwidth]{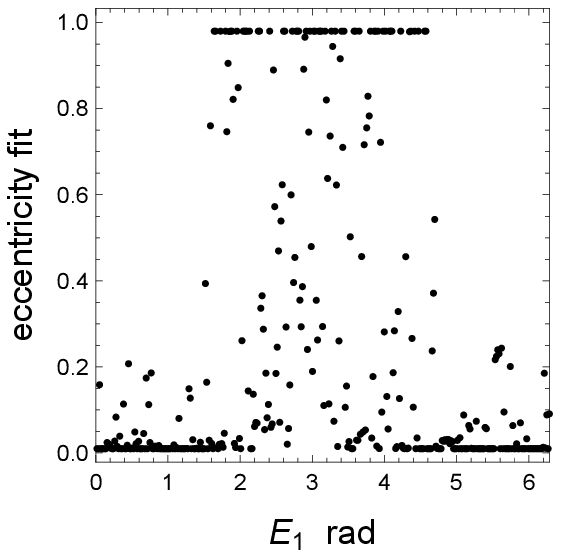}
\hspace{2pc}
\caption{Results of 360 two-point orbit solutions for HIP 69442 on a grid of fixed first-epoch eccentric
anomaly $E_1$, obtained with the Nelder-Mead optimization method. This data set also produced inconsistent
solutions with the fixed-eccentricity option.}
\label{fix2.fig}
\end{figure}

Fixed-$E_1$ grid solutions applied in \S\ref{impr.sec} to synthetic data can be used to gain some understanding
of the failed fits. Fig. \ref{fix2.fig} shows the Nelder-Mead solutions for one of the unsuccessful cases,
the binary system HIP 69442. According to \citet{izm}, the expected orbital period is 1355 years and the eccentricity
is 0.81. Although the merit function (left plot) has a well-defined minimum at around $E_1=4$, the corresponding
eccentricity estimates (right plot) do not show any clumping, with many fits bumping into the upper or lower constraints.
The period estimates are all bound to the upper limit. The four condition equations (\ref{eq.eq}) in this case are
weighted with weights inversely proportional to the combined formal errors of the corresponding data from Hipparcos
and Gaia. Nominally, the squared norm of the residual vector is expected to be equal to the mean of $\chi^2_4$, which is 4.
We can see that in this case the range of merit function values is quite narrow, with the smallest values much greater than
the expected value. This implies that the fitting model is inconsistent with the present data. To summarize, a failed
solution can be spotted by 1) lack of consistent eccentricity estimates around the global minimum; 2) best solutions
bumping into the hard constraints on eccentricity or period; 3) smallest merit function values being too large compared
to the expectation.

\section{Summary and discussion}
The technique of two-epoch orbit evaluation is viable in principle, but the practical difficulties arising
in working with real space astrometry data indicate some deficiencies. I believe the main problems are
the relatively modest precision of Hipparcos data for resolved double stars, especially the proper motions, and the intrinsic
poor condition of the full-rank optimization problems. Hipparcos could confidently
measure only stars brighter than magnitude $\sim 9$, but the situation was further exacerbated by the
blending of signal from the components at separations up to 20 arcsec. As a result, the magnitude difference
had to be small for a reliable determination of the secondary's proper motion to be made. The output of the Hipparcos data
reduction process was sensitive to initial assumptions about the components. Incorrect initial assumptions
sometimes led to gross errors even for apparently single stars \citep{fab}. It is not known if the formal
errors of astrometry are realistic for DMSA, so that our selection of S$/$N $>3$ cases does not let statistical
flukes through. 

On the other hand, Gaia has a vastly better dynamical range and better resolved tracking patches. Millions of
resolved double stars are present in DR2 with individual proper motions, despite the absence of a dedicated
double star pipeline. Being overall of much better quality, the Gaia part of the input data may still be subject
to physical perturbations and errors. Hierarchical multiple systems are common, and the proper motion of unresolved
tight pairs is perturbed by astrometric binarity \citep{wil, maka}. For Gaia, even large planets around nearby
stars can be detectable from proper motion perturbations \citep{kama}. Such complicated cases require models with
larger numbers of unknowns, for which the two-epoch technique provides bogus results. Finally, a fraction of
the Hipparcos-Gaia sample may be optical doubles. The proper motion differences in this case reflect only the
chance relative motion of two independent stars and random errors in the two catalogs.

It is more difficult to tackle the intrinsic poor condition of the two-epoch orbit solution,
but there may be possible ways to improve this as well. The method is expandable with more position measurements.
For example, if reliable angular separation and position angle measurements are available from, e.g.,
ground-based speckle interferometry or long-baseline optical interferometry, they can be incorporated
into the main matrix equation \ref{eq.eq} by adding two lines in matrix $\bf{S}$ with one additional unknown
$E_3$, two new data points, and one additional hard constraint on $E_3$ from the Kepler's equation. The method
then becomes a three-epoch estimation. Furthermore, spectroscopic radial velocity measurements of both
companions can be utilized as well, either as additional conditions in the matrix equation, or simply as
additional constraints in the optimization procedure.

Finally, the future plans for astrometric space missions open up excellent prospects for the proposed method.
The current Gaia mission could be used as the first epoch for millions of visual double stars, while a Gaia-NIR
successor \citep{hob, mca} would provide the second point separated by 20--25 years in time with an equal
or higher accuracy. The current techniques of ground-based orbit determination (speckle imaging, long-baseline
optical interferometry) will prove insufficient to process that great a number of targets. As we move to distances
beyond $\sim 100$ pc, angular separations become smaller, proper motion differences caused by orbital motion
become smaller too. The sample of characterized wide binaries can be greatly enlarged by an ultra-precise astrometric 
instrument, such as the anticipated performance of the narrow-angle, differential astrometry mission Theia \citep{the}.

\label{dis.sec}
\section*{Acknowledgments}
This research has made use of the VizieR catalogue access tool, CDS,
Strasbourg, France. The original description of the VizieR service was
published by \citet{och}.

\section*{Appendix}
A compilation of 1295 resolved double star pairs with individual component measurements in Hipparcos and Gaia DR2
(i.e., components' positions, parallaxes, and proper motions) is published as an online-only table. It includes
2590 rows divided into 1295 pairs. Each pair of rows include data for the A component (nominally, the brighter one
in Hipparcos) in the first row followed by a row of B component data. Each row contains 26 data fields. The contents of
this table are described in Table~1.

\begin{deluxetable}{lll}
\tablecaption{Resolved double star pairs with individual component measurements in Hipparcos and Gaia DR2}
\label{cat.tab}
\tablehead{
\colhead{Number} & \colhead{Units} & \colhead{Description}
}
\startdata
1 & --- & Hipparcos identification number \\
2 & mag & $H_p$ magnitude\\
3 & deg & Hipparcos right ascension in degrees\tablenotemark{a}\\
4 & deg & Hipparcos declination in degrees\tablenotemark{a} \\
5 & mas & Hipparcos parallax \\
6 & mas yr$^{-1}$ & Hipparcos proper motion in right ascension times $\cos\delta$\tablenotemark{b} \\
7 & mas yr$^{-1}$ & Hipparcos proper motion in declination\tablenotemark{b} \\
8 & mas & formal error of Hipparcos right ascension  times $\cos\delta$\tablenotemark{a}\\
9 & mas & formal error of Hipparcos declination\tablenotemark{a}\\
10 & mas & formal error of Hipparcos parallax\\
11 & mas yr$^{-1}$ & formal error of Hipparcos proper motion in right ascension  times $\cos\delta$\tablenotemark{b}\\
12 & mas yr$^{-1}$ & formal error of Hipparcos proper motion in declination\tablenotemark{b}\\
13 & --- & Gaia source id \\
14 & deg & Gaia right ascension in degrees\tablenotemark{a}\\
15 & deg & Gaia declination in degrees\tablenotemark{a} \\
16 & mas & Gaia parallax \\
17 & mas yr$^{-1}$ & Gaia proper motion in right ascension times $\cos\delta$\tablenotemark{b} \\
18 & mas yr$^{-1}$ & Gaia proper motion in declination\tablenotemark{b} \\
19 & mas & formal error of Gaia right ascension  times $\cos\delta$\tablenotemark{a}\\
20 & mas & formal error of Gaia declination\tablenotemark{a}\\
21 & mas & formal error of Gaia parallax\\
22 & mas yr$^{-1}$ & formal error of Gaia proper motion in right ascension  times $\cos\delta$\tablenotemark{b}\\
23 & mas yr$^{-1}$ & formal error of Gaia proper motion in declination\tablenotemark{b}\\
24 & mag & Gaia $G$ magnitude\\
25 & mag & Gaia $B_p$ magnitude\tablenotemark{c}\\
26 & mag & Gaia $R_p$ magnitude\tablenotemark{c}\\
\enddata
\tablecomments{Table 1 is published in its entirety in the electronic 
edition of the {\it Astronomical Journal}.  A portion is shown here 
for guidance regarding its form and content. Hipparcos astrometric data refer
to the mean epoch of $1991.25$. Gaia DR2 data refer to the mean epoch of $2015.5$.}
\tablenotetext{a}{ Hipparcos and Gaia J2000 equatorial coordinates in degrees are given for A-components
rows, which are odd-numbered in the Table (i.e., $1,3,5\ldots$). The B-component rows, which are even-numbered,
contain the coordinate differences ``B-component minus A-component" in mas instead. The corresponding formal
errors also refer to absolute coordinates for A-components and relative coordinates in B-component rows.}
\tablenotetext{b}{ Hipparcos and Gaia proper motions are given for A-components
rows, which are odd-numbered in the Table (i.e., $1,3,5\ldots$). The B-component rows, which are even-numbered,
contain the proper motion differences ``B-component minus A-component" instead. The corresponding formal
errors also refer to absolute proper motions for A-components and relative proper motions in B-component rows.}
\tablenotetext{c}{ If not available, $-999$ is given.}

\end{deluxetable}

\label{lastpage}

\end{document}